\documentclass[aps,prb,twocolumn,floatfix,nofootinbib,superscriptaddress,longbibliography,nobibnotes]{revtex4-2}
\usepackage[utf8]{inputenc}
\usepackage{bm}
\usepackage{physics}
\usepackage{xcolor}
\usepackage[final,colorlinks,bookmarks=true,citecolor=red,linkcolor=red,urlcolor=blue]{hyperref}
\usepackage{graphicx}
\usepackage{mathbbol}

\immediate\write18{latexdiff prb_v2/main_v2.tex main.tex > output_diff.tex}

\begin{document}

\title{Charge-4\textit{e} superconductivity in a Hubbard model}

\author{Martina~O.\,Soldini}
\affiliation{Department of Physics, University of Zurich, Winterthurerstrasse 190, CH-8057 Z\"{u}rich, Switzerland}

\author{Mark~H.\,Fischer}
\affiliation{Department of Physics, University of Zurich, Winterthurerstrasse 190, CH-8057 Z\"{u}rich, Switzerland}

\author{Titus~Neupert}
\affiliation{Department of Physics, University of Zurich, Winterthurerstrasse 190, CH-8057 Z\"{u}rich, Switzerland}

\date{\today}

\begin{abstract}
A phase of matter in which fermion quartets form a superconducting condensate, rather than the paradigmatic Cooper pairs, is a recurrent subject of experimental and theoretical studies. 
However, a comprehensive microscopic understanding of charge-4$e$ superconductivity as a quantum phase is lacking. 
Here, we study a two-orbital tight-binding model with attractive Hubbard-type interactions. 
Such a model naturally provides the Bose-Einstein condensate as a limit for electron quartets and supports charge-4$e$ superconductivity, as we show by mapping it to a spin-1/2 chain in this perturbative limit.
Using density matrix renormalization group calculations for the one-dimensional case, we further establish that the ground state is indeed a superfluid phase of 4$e$ charge carriers and that this phase can be stabilized well beyond the perturbative regime. 
Importantly, we demonstrate that 4$e$ condensation dominates over 2$e$ condensation even for nearly decoupled orbitals, which is a more likely scenario in electronic materials. 
Our model paves the way for both experimental and theoretical exploration of 4$e$ superconductivity and provides a natural starting point for future studies beyond one dimension or more intricate 4$e$ states.
\end{abstract}

\maketitle

\section{Introduction}
Electron pairing is the generic instability of a Fermi liquid in the presence of attractive interactions~\cite{PhysRev.104.1189}. This result prompts the question whether superconductivity based on four-electron condensates could exist in physical systems or whether such a phase will always be pre-empted by two-electron superconductivity. Experimentally, definite evidence for charge-4\textit{e} superconductivity is to date missing. Yet, unusual flux quantization in kagome metals were interpreted as signatures of 4$e$ superconductivity~\cite{PhysRevX.14.021025, pan2022frustrated, Zhou_2022}, and  the breaking of time-reversal symmetry (TRS) above $T_{\rm c}$ in some iron-based superconductors was argued to be compatible with fermion quadrupling~\cite{2021NatPh..17.1254G,PhysRevB.107.064501,PhysRevB.105.214520}, although without establishing phase coherence. 

Still, several theoretical studies proposed specific scenarios or broader mechanisms that would enable 4\textit{e}-superconducting phases. These include the thermal fluctuation regime above a pair-density-wave phase, in which 4\textit{e}-superconductivity appears as a vestigial order~\cite{PhysRevLett.95.266404,2009NatPhysKivelson,PhysRevB.79.064515,reviewPDW,PhysRevLett.127.227001,volovik2023fermionic}, doping an antiferromagnet~\cite{PhysRevB.42.6523}, the coupling of two superconducting condensates~\cite{PhysRevB.82.134511,PhysRevLett.127.047001}, by frustrating the superconductivity in 45-degree twisted bilayer cuprates~\cite{Charge4etwisted}, through condensation of Skyrmions in a quantum spin-Hall phase~\cite{PhysRevB.85.245123}.
Other attempts have tackled models built as mean-field analogues of  Bardeen–Cooper–Schrieffer (BCS) theory for a $4e$ order parameter~\cite{PhysRevB.95.241103,PhysRevB.106.094508}, or by studying a $4e$ analog of a BCS wavefunction~\cite{https://doi.org/10.48550/arxiv.2209.13905} corresponding to a highly non-local Hamiltonian. Despite this plethora of proposals, a microscopic understanding of the quantum nature of 4$e$ superconductivity is missing. 

%A major road block is the lack of a well-behaved microscopic model system amenable to numerical techniques.
A body of work that is motivated by experiments with ultracold atoms studied SU($N$) Hubbard models in one dimension (1D)~\cite{PhysRevLett.95.240402,PhysRevA.77.013624,PhysRevB.75.100503,Roux_2008,Yoshida_2021,Yoshida_2022}. These systems host quite generically so-called molecular superfluid phases, which are in essence charge-$N$e superconductors. For condensed matter systems, the phase robustness away from the SU($N$) symmetric limit~\cite{PhysRevA.77.013624} and  realizations in other dimensions are of great interest, but under-explored. 

In this work, we study charge-$4e$ superconductivity in an attractive Hubbard-type model, which is local and appeals through its simplicity. The attractive Hubbard model has been a vital theoretical test bed for establishing superconductivity beyond the BCS mean-field approximation~\cite{RevModPhys.62.113}. As such, it allows us to explore the cross-over from BCS to a Bose-Einstein-condensate (BEC) regime~\cite{PhysRevB.24.4018,PhysRevB.26.3915,Nozieres1985} and the competition between charge density wave (CDW) order and singlet superconductivity.
The BEC limit of the Hubbard model is a particularly natural starting point: For large enough attractive interactions and no hopping between the sites, fermions are bound into pairs at each lattice site. Adding hopping between the sites as a perturbation then allows the pairs to move and establish phase coherence at low enough temperatures for appropriate fillings~\cite{PhysRevA.79.033620}. This physics is well described through mapping the effective low energy model to a hard-core boson or spin model~\cite{Singer1998}. 
Here, we generalize the idea of the BEC limit as a starting point by considering two orbitals per site and choosing the interactions between the states such as to form four-electron bound states. While the model realizes an SU(4) Hubbard model when inter- and intra-orbital interactions are equal, we focus on the case of inter-orbital interactions weaker than intra-orbital interactions, a scenario more likely to occur in electronic materials. 

Note that one-dimensional (1D)  superconductors are gapless states, for which  correlations decay algebraically as a function of distance. Specifically, we define a 1D 4\textit{e} superconductor as a state in which four-body correlations decay algebraically, while two- and one-body correlations decay exponentially. As a consequence of the charge carriers being quartets, in other words $e^*=4e$, this phase should respond to flux insertion with a flux quantization of a quarter of the flux quantum, $\Phi_0/4 = h/(4e)$. In this work, we explore the stability and properties of a charge-4$e$ phase in the thermodynamic limit, using the infinite 
density matrix renormalization group (iDMRG) algorithm~\cite{PhysRevLett.69.2863,PhysRevB.48.10345,RevModPhys.77.259,mcculloch2008infinite}.

\section{Model}
\begin{figure}
    \centering
\includegraphics{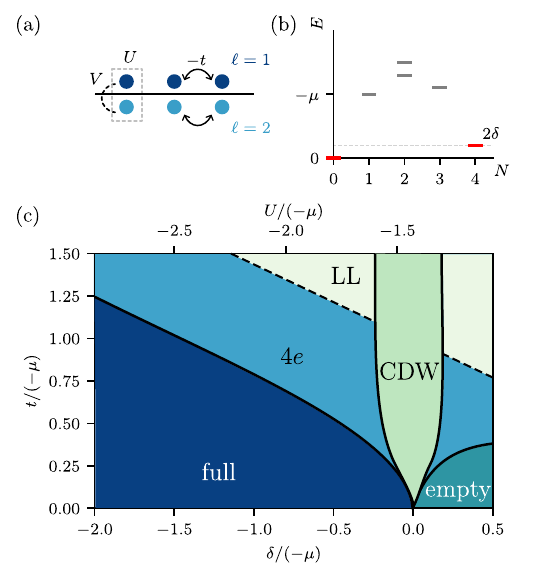}
    \caption{The charge-$4e$ model introduced in Eq.~\eqref{eq:1D charge4e model}: (a) Real-space schematic with a unit cell highlighted by the gray dashed rectangle. (b) Energy levels as a function of particle number $N$ for the single-site Hamiltonian, Eq.~\eqref{eq:single site H} ($U=-0.5$, $V=-0.7$). (c) Schematic phase diagram as a function of $\delta$ and $t$, evaluated at constant $V/U=0.2$. The empty and full phases correspond to regions where the ground state is the empty or full state, respectively, the CDW phase corresponds to a $Q=\pi$ CDW state, and the $4e$ phase indicates the charge-$4e$ superconducting region of the phase diagram. The dashed line indicates the onset of a Luttinger liquid phase, labeled by LL, which breaks down the $4e$ superconductivity and suppresses the CDW. The phase boundaries are traced from the iDMRG data obtained at bond dimension $\chi=500$ [App.~\ref{app:iDMRG phase diagram}].}
    \label{fig:charge4e model}
\end{figure}

 It is instructive to first construct the BEC limit for the 4\textit{e}-superconducting phase. Working in the grand-canonical ensemble, we consider a lattice system with---for now---no tunneling between unit cells, and a (near) degeneracy between ground states, which differ in filling by four electrons. To realize this situation, we consider a 1D chain with each site containing two spinful orbitals, labeled by $\ell \in \{1, 2\}$ and spin index $\sigma \in \{\uparrow, \downarrow\}$ [Fig.~\ref{fig:charge4e model}(a)]. Hence, every site has four single-particle states, labeled by orbital $\ell$ and spin $\sigma$, and we further denote by $\hat{c}^{\dagger}_{i, \ell, \sigma}$ ($\hat{c}_{i, \ell, \sigma}$) the creation (annihilation) operator for such an electronic state at the $i$th site.
The corresponding density operator is $\hat{n}_{i, \ell, \sigma}=\hat{c}^{\dagger}_{i, \ell, \sigma} \hat{c}_{i, \ell, \sigma}$, and we define  $\hat{n}_{i, \ell} = (\hat{n}_{i, \ell, \uparrow} + \hat{n}_{i, \ell, \downarrow})$. The onsite Hamiltonian for the $i$-th site reads
\begin{equation}\label{eq:single site H}
    \hat{H}_{i} = - \mu \sum_{\ell, \sigma} \hat{n}_{i, \ell, \sigma} + U \sum_{\ell} \hat{n}_{i, \ell, \uparrow} \hat{n}_{i, \ell, \downarrow} + V \hat{n}_{i, 1} \hat{n}_{i, 2},
\end{equation}
where $\mu $ is the chemical potential, and $U$ and $V$ are the  Hubbard interaction parameters within the same and between different orbitals, respectively. In the following, we focus on the regime  $\mu,\, U,\, V<0$ and express all the parameters in units of $(-\mu)$ for compactness of notation.
The onsite Hamiltonian, Eq.~\eqref{eq:single site H}, is diagonal in the occupation number basis, and the four-fold occupied state has energy
\begin{equation}
    E_4 = 2\delta, \quad \delta \equiv 2 + U + 2 V.
\end{equation}
By tuning $\delta \ll 1$ and $U, \, V$ appropriately, the low energy subspace of the single-unit-cell Fock space comprises the empty ($N=0$) and four-fold occupied ($N=4$) states only, while the states with particle numbers $N=1, 2, 3$ lie at higher energy [Fig.~\ref{fig:charge4e model}(b)]. Note that this energy hierarchy can only be reached if $V \neq 0$. At $V=0$ the two orbitals are fully decoupled, and $N=2$ states become part of the low-energy subspace with the $N=4$ state. 

For a finite lattice of size $L$, the (near) degeneracy of zero- and four-electron states at every site leads to an extensive (near) ground-state degeneracy of states with $N=4n$ ($n=0, 1, \cdots L$) particles. In a thermodynamically large lattice, the (near) degeneracy occurs at all fillings $0<\nu<1$~\footnote{We define $\nu:=N/4L$ before the thermodynamic limit is taken.}.
The extensive ground-state degeneracy alone is not sufficient to reach a superconducting state. Thus, we connect neighboring sites by spin- and orbital-conserving tunneling terms, leading to the final Hamiltonian
\begin{equation}\label{eq:1D charge4e model}
    \hat{H} = \sum_{i=1}^L \hat{H}_{i} - t \sum_{i=1}^L \sum_{\ell, \sigma} (\hat{c}^{\dagger}_{i, \ell, \sigma} \hat{c}_{i+1, \ell, \sigma} + \text{H.c.}).
\end{equation}
By introducing tunneling between the unit cells, the degeneracy is lifted for finite $L$ and may, in the thermodynamic limit, result in a phase-coherent superconducting condensate whose constituents are electronic quartets~\cite{GREITER2005217}.

Figure.~\ref{fig:charge4e model}(c) shows the ground state phase diagram of Hamiltonian~\eqref{eq:1D charge4e model} as obtained from a two-site iDMRG algorithm [App.~\ref{app:iDMRG parameters}] for the case of $V/U=0.2$ [App.~\ref{app:iDMRG phase diagram}]. Importantly, with this approach, we work in the grand-canonical ensemble, directly in the thermodynamic limit. The phase diagram features $4e$ superconductivity in a large part of the parameter space,~\footnote{
At $V=0$, the two orbitals $\ell=1,2$ of each site are completely decoupled. In this limit, the 
system realizes two independent copies of a 1D Hubbard chain, one formed by the $\ell=1$ orbitals and the other by the $\ell=2$ orbitals. The attractive fermionic Hubbard chain is known to host a charge-$2e$ superconducting ground state for certain range of filling~\cite{PhysRevLett.20.1445,1979AdPhy..28..201S}[see also App.~\ref{app:charge2e model}], while the SU(4) symmetric case of $V=U$ was studied in Refs.~\cite{PhysRevLett.95.240402,PhysRevA.77.013624,PhysRevB.75.100503,Roux_2008}.}, separated by a ($Q\!=\!\pi$) CDW with filling $\nu=1/2$. While for zero hopping the system is either empty or full, finite hopping leads to a non-trivial filling and can introduce coherence between sites. At large values of $t$, where a transition to a Luttinger liquid phase occurs, the 4e superconducting phase is suppressed as indicated by the dashed line in Fig.~\ref{fig:charge4e model}(c). Using iDMRG, it is hard to reliably locate this transition between two gapless regimes.
Yet, the numerical evidence we discuss below, in Sec.~\ref{sec:Transfer matrix} and~\ref{sec:corr and susc}, shows the extent of charge-$4e$ superconductivity well beyond the perturbative regime in a substantial portion of the phase diagram. 
Note that the regions of $t\ll1$ and $\delta \ll 1$ of the phase diagram can be qualitatively understood through a mapping of the low-energy limit to an effective spin-1/2 model, which we discuss in the following. 

\section{Low energy effective model}\label{sec:Low energy effective model}
To see that the Hamiltonian~\eqref{eq:1D charge4e model} realizes the quartetting phase, we consider a low-energy effective model in the limit of $t\ll1$ and $\delta\ll1$.
This regime motivates a mapping of the Hamiltonian in Eq.~\eqref{eq:1D charge4e model} to a spin-1/2 chain, where the two spin states at every site correspond to the empty and four-electron state, namely the quartet of our model, of the fermionic sites.  As compared to the $2e$ case, the hopping of a quartet is a fourth-order process in $t$, competing with second-order processes, which makes the conclusion about the emergence of a phase-coherent condensate less immediate. Although the effective model has some degree of complexity due to fourth-order terms in perturbation theory, the result simplifies to an XXZ chain with external magnetic field and next-to-nearest neighbor coupling. We indicate the spin-1/2 operator at site $i$ by $\hat{S}^{j}_i = \sigma^j_i/2$ ($j=x, y, z$), with $\sigma^j$ the Pauli matrices. The low-energy effective spin Hamiltonian reads [App.~\ref{app:effective spin model}]
\begin{subequations}\label{eq:spin eff model}
\begin{equation}
\begin{split}
    \hat{H}_{\mathrm{spin}} =& \sum_{i} J (\hat{S}^x_i \hat{S}^x_{i+1} + \hat{S}^y_i \hat{S}^y_{i+1} + \Delta \hat{S}^z_i \hat{S}^z_{i+1})\\
    &+ h \sum_i \hat{S}^z_{i} + K \sum_i \hat{S}^z_i \hat{S}^z_{i+2},
\end{split}
\end{equation}
and, to leading order, the coefficients depend on $t$ and $\delta$ as
\begin{equation}
    \begin{split}
        J \sim t^4, \quad
      J \Delta  \sim  t^2, \quad
      K \sim t^4, \quad
      h  = 2\delta.
     \end{split}
\end{equation}
\end{subequations}
Their exact form is derived in App.~\ref{app:effective spin model}.
Without next-to-nearest-neighbor coupling, meaning $K=0$, this model realizes the paradigmatic XXZ model, which is known to host a paramagnetic gapless superfluid phase~\cite{Franchini2016}. A non-zero value of $K$ does not introduce frustration in the model, therefore it does not disrupt the underlying phases. The model in Eq.~\eqref{eq:spin eff model} can be equivalently expressed as a 1D attractive Hubbard model for hardcore bosons, as shown in App.~\ref{app:effective spin model}, which is also known to host a superfluid phase~\cite{PhysRevB.105.134502}. 

\begin{figure}[t!]
    \centering
    \includegraphics{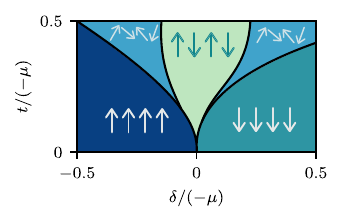}
    \caption{\textbf{Effective spin model.} Phase diagram of the effective spin model of Eq.~\eqref{eq:spin eff model} as a function of $\delta$, $t$ and fixed $V/U=0.8$. The phase boundaries are traced based on the spin polarization, its variance, and spin correlations as obtained from iDMRG [App.~\ref{app:effective spin model}]. The different spin phases are labeled by a pictorial representation of the spin configuration.}
    \label{fig:spin eff model}
\end{figure}

We compute the ground state of the effective model in Eq.~\eqref{eq:spin eff model} within two-site iDMRG, and draw the phase diagram as a function of the perturbative parameters $t$ and $\delta$, at fixed $V/U=0.8$ [Fig.~\ref{fig:spin eff model}]. For the low-energy effective model to be valid, we need $V/U$ to be of order 1, while the perturbation theory breaks down for $V/U \ll 1$.
At small $t/\abs{\delta}$, the ground state is ferromagnetic, with opposite polarization depending on the sign of $\delta$. At large $t/\abs{\delta}$, the system orders antiferromagnetically, and finally, for values of $t$ comparable to $\abs{\delta}$ ($t/\abs{\delta} \sim 1$), the ground state is in a superfluid gapless phase. In the language of the original fermionic model, the ferromagnetic phases with opposite polarization correspond to a completely empty or completely filled chain, the antiferromagnetic phase maps to a $Q=\pi$ CDW order, where sites are alternatively empty or four-fold occupied, and the gapless superfluid phase translates into a charge-$4e$ superconducting gapless phase.
Below, we use iDMRG in the original fermionic model to characterize the non-trivial regions of this phase diagram further.

\section{Numerical results}

We now come back to and study the fermionic Hamiltonian, Eq.~\eqref{eq:1D charge4e model}, beyond the aforementioned perturbative mapping to a spin-1/2 chain, where the existence of 4e superconductivity is rigorously established. In this section, we consider individual points in parameter space lying well within the phases of interest to explore their specific signatures in terms of transfer matrix properties, correlations and susceptibility. We consider iDMRG results with different values of bond dimension $\chi$ to extrapolate the infinite long-range distance behavior of gapless phases, which would require $\chi\rightarrow\infty$ for a full description. As the full and empty lattice states are simply gapped product states, we focus on the comparison between the 4$e$-superconducting phase, the CDW phase, and the 2$e$ superconductivity, which is restored in the limit $V=0$. For completeness, we show the data for the Luttinger liquid phase in App.~\ref{app:Luttinger Liquid}.

Based on the insights of this Section, we draw the phase boundaries in~Fig.~\ref{fig:charge4e model}(c) by considering iDMRG data at fixed bond dimension. The boundaries are then obtained by considering the value of correlations at finite distance, CDW order parameter, and filling of the iDMRG ground state [App.~\ref{app:iDMRG phase diagram}]. In App.~\ref{app:ED flux insertion}, we complement the iDMRG results with exact diagonalization data on a finite size chain of length $L=8$, where we show the evolution of the many body energy spectrum under magnetic flux insertion, which displays the expected $\Phi_0/4$ periodicity.

\subsection{Transfer matrix observables}\label{sec:Transfer matrix}
To characterize different phases, we analyze the transfer matrix $T$~\cite{doi:10.1146/annurev-conmatphys-031016-025507,10.21468/SciPostPhysLectNotes.7} constructed from the matrix-product (MPS) ground states obtained from iDMRG in the respective phases.
The transfer matrix is obtained by contracting the physical legs of the on-site MPS tensor obtained from iDMRG at fixed $\chi$ with its Hermitian conjugate. The resulting tensor has four virtual legs of dimension $\chi$, and can therefore be recast into the matrix $T$ of size $\chi^2 \times \chi^2$.
The transfer matrix can be decomposed in terms of its eigenvalues $\{\lambda_{i}\}_{i=1}^{\chi^2}$ and eigenvectors $\{\ket{v_i}\}_{i=1}^{\chi^2}$ (ordered as $\abs{\lambda_{1}}>\abs{\lambda_{2}} \geq \abs{\lambda_{2}} \geq \cdots \geq  \abs{\lambda_{{\chi^2}}} $)
\begin{equation}\label{eq:T decomposition}
    T = \sum_{i=1}^{\chi^2}  \lambda_{i} \ket{v_i} \bra{v_i}
\end{equation}
with the largest eigenvalue $\lambda_{1} = \max \{\lambda_{i}\}_{i=1}^{\chi^2}=1$, by construction and normalization constraints, and all the other eigenvalues $|\lambda_{i}|<1$ for $i\neq1$.

\begin{figure}[t]
    \centering
    \includegraphics{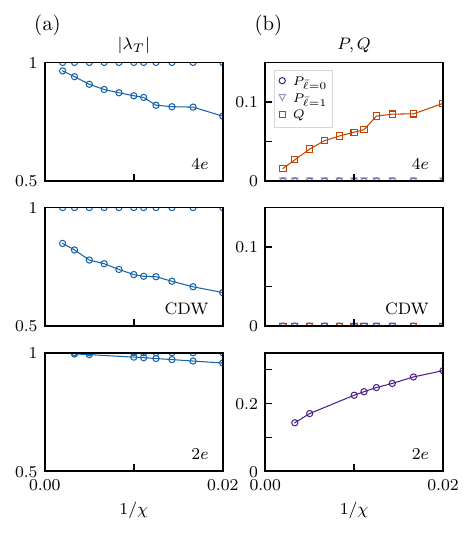}
\caption{Transfer matrix characterization of the charge-$4e$ (top), CDW (middle), and charge-$2e$ (bottom) phases, as obtained from the iDMRG ground state. (a) The two largest eigenvalues in absolute value of $T$ are shown as a function of inverse bond dimension $1/\chi$. (b) Transfer matrix eigenvector overlaps with the pair operators $P_{\bar{\ell}=0}$, $P_{\bar{\ell}=1}$, and $Q$ as a function of $1/\chi$. The $2e$ panels are evaluated on the charge-$2e$ model [App.~\ref{app:charge2e model}], hence only $P_{\bar{\ell}=0}$ is shown. The parameters for the panels are $(\delta,\, t, \, V/U) = (-0.5, \, 0.6,\, 0.2)$ for the $4e$, $(0, \, 1,\, 0.2)$ for the CDW, and $(-0.3, \, 0.5,\, 0)$ for $2e$ panels.}
    \label{fig:transfer matrix}
\end{figure}

This matrix encodes information on the decay of correlations in the ground state: If the subleading eigenvalue of $T$ is separated from 1 by a gap that persists in the limit $\chi\to \infty$, all correlations decay exponentially over large enough distances.
Another way to see this is to note that the correlation length $\xi$ of a state can be extracted from the second leading eigenvalue of $T$, $\lambda_{2}$, through the relation $\lambda_{2} = e^{-\frac{1}{\xi}}$~\cite{PhysRevX.8.041033}. Therefore, $\xi \xrightarrow{\chi \rightarrow \infty} \infty$ is equivalent to $\lambda_{2} \xrightarrow{\chi \rightarrow \infty} 1$.
Indeed, the subleading eigenvalue extrapolates to 1 in this limit in the putative charge-$2e$ and charge-$4e$ phases, while this is not the case in the CDW phase~[Fig.~\ref{fig:transfer matrix}(a)]. This allows in principle for algebraic decay of correlations in the former two phases, bearing in mind that MPS at finite $\chi$ are not able to fully capture this behavior~\cite{PhysRevX.8.041033}.

To motivate our approach to analyze $T$, let us consider connected correlators of the following form, for a translationally invariant system~\cite{10.21468/SciPostPhysLectNotes.7,PhysRevX.8.041033}
\begin{equation}
    C_{O}(|i-j|) = \expval{\hat{O}^{\dagger}_i \hat{O}_j}_{\mathrm{GS}} -
    \expval{\hat{O}^{\dagger}_i}_{\mathrm{GS}}  \expval{\hat{O}_j}_{\mathrm{GS}},
\end{equation}
where $\hat{O}_i$ is an operator acting on site $i$.
These can be expressed as~\cite{10.21468/SciPostPhysLectNotes.7}
\begin{equation}
    C_{O}(|i-j|) = \tr [\hat{O}^{\dagger}_i T^{|i-j-1|} \hat{O}_j] -
    \expval{\hat{O}^{\dagger}_i}_{\mathrm{GS}}  \expval{\hat{O}_j}_{\mathrm{GS}}
\end{equation}
where $T$ is the transfer matrix. 
By inserting Eq.~\eqref{eq:T decomposition}, the connected correlator becomes
\begin{equation}
    C_{O}(|i-j|)  = \Big\langle\hat{O}^{\dagger}_i \left(\sum_{i=2}^{\chi^2} \lambda^{|i-j+1|}_{v_i}  \ket{v_i} \bra{v_i}\right) \hat{O}_j\Big\rangle_{\mathrm{GS}}.
\end{equation}
Therefore, maximizing the overlap between the second leading eigenvalue $\ket{v_2}$ and a generic operator $\hat{O}$, conveys which operator is characterized by the slowest exponential decay across the MPS ground state. This, for a gapless phase, would lead to algebraic decay in the limit of $\chi \rightarrow \infty$, combined with the correlation length divergence.

In practice, there is a finite number of symmetry in-equivalent operators in our model, therefore it is sufficient to evaluate the overlaps between $\ket{v_2}$ and a finite subset of operators to find the leading overlaps.
To this end, we introduce the pair and quartet operators
\begin{equation}
    \begin{split}
    \hat{P}_{r, \ell, \sigma, \bar{\ell}, S} &= \hat{c}_{r, \ell, \sigma} \hat{c}_{r, \ell+\bar{\ell}, \sigma + S}, \\ \hat{Q}_r &= \hat{c}_{r, 1, \uparrow}\hat{c}_{r, 1, \downarrow}
    \hat{c}_{r, 2, \uparrow}\hat{c}_{r, 2, \downarrow}.
    \end{split}
\end{equation}
Since the Hamiltonian in Eq.~\eqref{eq:1D charge4e model} conserves TRS, particle number, orbital and spin quantum numbers, it is sufficient to consider the subset $P_{r, \bar{\ell}} \equiv P_{r, 1, \uparrow, \bar{\ell}, S}$ out of all the two-body operators to have a minimal set of operators spanning all the symmetry in-equivalent two-particle operators.

The overlaps between the eigenstate of $T$ with subleading eigenvalue and the $\hat{P}_{\bar{\ell}}$ and $\hat{Q}$ operators are shown in Fig.~\ref{fig:transfer matrix}(b) as a function of $1/\chi$. We indicate the overlaps with the pair and quartet operators by $P_{\bar{\ell}}$ and $Q$, respectively.
While $P_{\bar{\ell}=0}$ dominates in the charge-$2e$ superconducting phase, $Q$ dominates in the charge-$4e$ phase, and no overlap has a substantial amplitude in the CDW phase, as compared to the two former phases. This indicates that the operators with slowest decay across large distances, ultimately becoming algebraic at $\chi \rightarrow \infty$, are the quartet operator in the charge-$4e$ phase and a pair operator in the charge-$2e$ phase. 

\subsection{Correlations and susceptibility}\label{sec:corr and susc}
\begin{figure*}[th]
    \centering
    \includegraphics{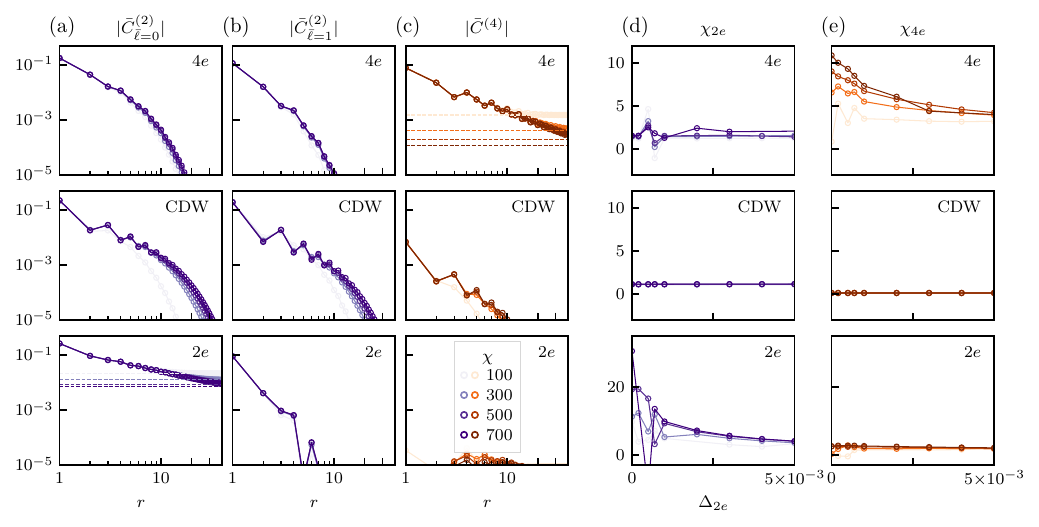}
    \caption{Characterization through correlations and susceptibility of the charge-$4e$ (top row), CDW (middle row), and charge-$2e$ (bottom row) phases, as obtained from iDMRG. (a)--(c) Log-log plot of the absolute value of the (a) $\bar{C}^{(2)}_{\bar{\ell}=0}$, (b) $\bar{C}^{(2)}_{\bar{\ell}=1}$ and (c) $\bar{C}^{(4)}$ defined in Eq.~\eqref{eq:corr with subtraction} as a function of distance $r$. The dashed horizontal lines indicate the value of $|\langle \hat{P}_{\bar{\ell}=0}\rangle|^2$, in panel (a) bottom row, and $|\langle\hat{Q}_{\,}\rangle|^2$, in panel (c) upper row, at different $\chi$, corresponding to the value at which the algebraically decaying correlations level off for the $2e$ and $4e$ case respectively~[App.~\ref{app:correlations}]. (d)-(e) Susceptibility to (d) charge-$4e$ and (e) charge-$2e$ perturbations, see Eq.~\eqref{eq:susceptibilities}. Throughout the figure the color opacity indicates the bond dimension $\chi$, while purple and orange distinguish $2e$ and $4e$ probes, respectively.
    The parameters for the panels are the same as the one considered in~Fig.~\ref{fig:transfer matrix}.}
    \label{fig:corr and suscep}
\end{figure*}
After having discussed the transfer matrix properties in different regions of the phase diagram, we analyze the correlations and susceptibility to a superconducting mean-field perturbation.

\paragraph{Correlations.}
We define the reduced two- and four-body correlation functions
\begin{equation}\label{eq:corr with subtraction}
\begin{split}
    \bar{C}^{(4)}(r) &= C^{(4)}(r) - [C^{(2)}_{\bar{\ell}=0}(r)]^2, \\
    \bar{C}^{(2)}(r) &= C^{(2)}_{\bar{\ell}=0} - [C^{(1)}(r)]^2,
\end{split}
\end{equation}
where
$C^{(2)}_{\bar{\ell}, S}(r) = \abs{ \expval{\hat{P}^{\dagger}_{0,\bar{\ell}} \hat{P}_{r, \bar{\ell}}}}$, 
$C^{(4)}(r)= \abs{\expval{\hat{Q}^{\dagger}_{0} \hat{Q}_{r}}}$, and $C^{(1)}_{\ell, \sigma}(r) = \abs{\expval{\hat{c}^{\dagger}_{0, \ell, \sigma} \hat{c}_{r, \ell, \sigma}}}$. Due to the symmetries of the Hamiltonian, all the $C^{(1)}_{\ell, \sigma}(r)\equiv C^{(1)}(r)$'s are equal. In Eq.~\eqref{eq:corr with subtraction} we already used the ensuing simplification by dropping the respective indices.
All the expectation values are computed over the ground state.

Figures~\ref{fig:corr and suscep}(a)--\ref{fig:corr and suscep}(c) show the correlations, Eq.~\eqref{eq:corr with subtraction}, evaluated on the ground state in three different phases, as a function of distance $r$ and different bond dimension $\chi$ of the MPS ansatz. In the $4e$-superconducting phase, the two-body correlations decay exponentially for any $\chi$, while the four-body correlations decay algebraically before leveling off at a distance that increases with $\chi$.
For the CDW case, all the correlations decay exponentially. At $V/U=0$, the charge-$2e$ superconducting phase has ${\bar{C}^{(2)}_{\bar{\ell}=0}}$ correlations following an algebraic decay, which then level off at large distances, while ${\bar{C}^{(2)}_{\bar{\ell}=1}}$ and $\bar{C}^{(4)}$ are exponentially suppressed. 

The constant behavior at large $r$ in the $4e$ and $2e$ cases is an artifact of the finite $\chi$~\cite{PhysRevB.105.134502}, and an intrinsic limitation of the MPS ansatz to capture gapless systems~\cite{PhysRevX.8.041033}. The constant value at which correlations level off can be directly related to the local expectation values of particle nonconserving operators, namely
\begin{equation}\label{eq:corr large r}
   \lim_{r\rightarrow\infty} C^{(2)}_{\bar{\ell}}(r) = |\langle \hat{P}_{0, \bar{\ell}} \rangle|^2, \quad \lim_{r\rightarrow\infty} C^{(4)}(r) = |\langle \hat{Q}_{0} \rangle|^2.
\end{equation}
The squared expectation values as in the right hand sides in Eq.~\eqref{eq:corr large r}, are indicated by the dashed lines in the relevant panels of Fig.~\ref{fig:corr and suscep}, and they coincide with the value at which the algebraically decaying correlations level off at large distances.  From the above arguments it follows that the connected correlations, namely ${\langle P^{\dagger}_{0, \bar{\ell}=0} P_{r,\bar{\ell}=0}\rangle - \langle P^{\dagger}_{r, \bar{\ell}=0}
\rangle \langle P^{\,}_{r, \bar{\ell}=0}}\rangle$ and ${\langle Q^{\dagger}_0 Q_r \rangle - \langle Q^{\dagger}_0\rangle \langle Q^{\,}_0}\rangle$, exponentially decay at large distances instead of leveling off, and this is shown in~App.~\ref{app:correlations}. 

The observations above can be explained by noting that even though the Hamiltonian in Eq.~\eqref{eq:1D charge4e model} does not break particle-number conservation, the ground state obtained from the iDMRG algorithm at finite $\chi$ acquires a nonzero expectation value of the pair operator $\hat{P}_{\bar{\ell}=0}$ and quartet operator $\hat{Q}$ in the $2e$ and $4e$ phases, respectively. This is allowed because particle number conservation is not imposed in the iDMRG simulations~[App.~\ref{app:iDMRG parameters}]. The true ground state, which would be obtained at $\chi \rightarrow \infty$, in general does not mix different particle number sectors, and therefore we expect the pair and quartet expectation values to go to zero as $\chi$ increases, and similarly for the particle number fluctuations. The former are shown in App.~\ref{app:correlations}.
For the latter, we carry out a numerical analysis in App.~\ref{app:variance N}, where we consider the quantity
\begin{equation}\label{eq:variance N finite L}
    \text{var}(\hat{N}_L)/L = (\langle \hat{N}_L^2\rangle - \langle \hat{N}_L \rangle^2)/L,
\end{equation}
with $\hat{N}_L$ the particle number operator on a system of finite size $L$, and show that Eq.~\eqref{eq:variance N finite L} extrapolated to $L\rightarrow\infty$ approaches zero for increasing bond dimension $\chi$.
For completeness, the bare correlations (without subtraction) are shown in App.~\ref{app:correlations}.

\paragraph{Susceptibilities.}
Figures~\ref{fig:corr and suscep}(d)--\ref{fig:corr and suscep}(e) show the susceptibilities to the charge-$4e$ and charge-$2e$ superconducting perturbations for the same choices of model parameters considered in Fig.~\ref{fig:transfer matrix}. In particular, we study the susceptibilities to mean-field perturbations of charge-2\textit{e} and -4\textit{e} type
with perturbations of the form 
\begin{subequations}\label{eq:mean field perturbs}
\begin{equation}
\delta \hat{H}_{2e} = \Delta_{2e} \sum_{i, \ell} (\hat{c}^\dagger_{i,\ell, \uparrow}\hat{c}^\dagger_{i,\ell, \downarrow} + \text{H.c.})
\end{equation} 
and
\begin{equation}
\delta \hat{H}_{4e} = \Delta_{4e} \sum_{i} (\hat{Q}^\dagger_{i} + \text{H.c.}),
\end{equation} 
\end{subequations}
respectively.
These perturbations lead to non-vanishing expectation values 
$|\langle \hat{P}^\dagger_{\bar{\ell}=0}\rangle|$ and $|\langle\hat{Q}^\dagger\rangle|$ with respect to the ground state of the perturbed Hamiltonian, where we dropped the index $r$ due to translational symmetry.
Finally, we obtain the susceptibilities
\begin{equation}\label{eq:susceptibilities}
    \chi_{2e} = \left. \pdv{\abs{\langle\hat{P}^\dagger_{\bar{\ell}=0}\rangle}}{\Delta_{2e}}\right\rvert_{\Delta_{4e}=0}, \quad
    \chi_{4e} = \left. \pdv{\abs{\langle\hat{Q}^{\dagger}\rangle}}{\Delta_{4e}}\right\rvert_{\Delta_{2e}=0}.
\end{equation}
While the exact order parameters for the charge-2$e$ and charge-4$e$ superconductors are in principle not known, we expect a non-zero overlap with $\hat{P}_{\bar{\ell}=0}$ and $\hat{Q}^{\dagger}$, respectively. 
Importantly, for a choice of parameters falling in the charge-$4e$ phase $\chi_{4e}$ shows a peak for $\Delta_{4e}\rightarrow0$, while the charge-$2e$ phase has a peak in $\chi_{2e}$, and the susceptibilities are constant in the CDW phase. For completeness, the expectation values of $\hat{P}_0$ and $\hat{Q}$ are shown in App.~\ref{app:susceptibility}.

\section{Stability of the charge-4$e$ phase} 
\begin{figure}[t]
    \centering
    \includegraphics{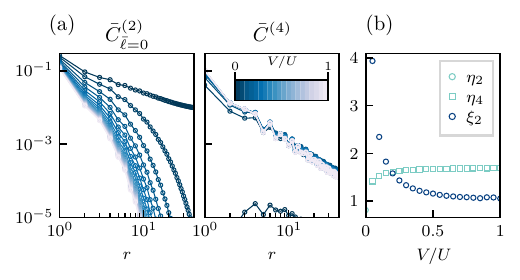}
    \caption{(a) Log-log plot of the correlations defined in Eq.~\eqref{eq:corr with subtraction} for varying $V/U\in [0, 1]$, constant $\delta=-0.3$ and $t=0.5$, and at fixed bond dimension $\chi=500$. (b) Exponents extracted from the curves in (a), as defined in Eq.~\eqref{eq:exponents correlations}. At $V/U=0$ only $\eta_2$ is defined.}
    \label{fig:V_U}
\end{figure}

So far, we have focused on a few values of $V/U$, specifically $V/U=0.2$ and $V=0$, to establish that they realize, respectively, charge-$4e$ and charge-$2e$ superconducting phases.
We now address the interpolating regime $0\leq V \leq U$, while keeping $\delta$ constant. Figure~\ref{fig:V_U}(a) shows the correlations introduced in Eq.~\eqref{eq:corr with subtraction}, and we drop $\bar{C}^{(2)}_{\bar{\ell}=1}$ as it remains exponentially decaying both in the 2e and 4e superconducting phases. While $\bar{C}^{(4)}$ remains algebraic for any $V<0$, $\bar{C}^{(2)}_{\bar{\ell}=0}$ switches from an exponential decay to algebraic at $V/U=0$, suggesting a $4e$ phase for any nonzero $V$.
We parametrize the long-distance behavior of the correlations as
\begin{equation}\label{eq:exponents correlations}
    \bar{C}^{(i)}(r)\sim r^{-\eta_i}, \quad \bar{C}^{(i)}(r)\sim e^{-\frac{r}{\xi_i}},
\end{equation}
for algebraic and exponential decay, respectively. While the MPS ground state cannot fully capture an algebraic decay~\cite{PhysRevX.8.041033}, we extrapolate the value of $\eta_i$'s by considering the correlations up to a finite distance, after which they become constant.
In Fig.~\ref{fig:V_U}(b), the extracted $\xi_2, \, \eta_4$ for $V/U>0$, and $\eta_2$ for $V/U=0$ are shown.
Remarkably, the charge-$4e$ phase remains stable over a large parameter range away from $V=U$, the SU(4) limit where quartetting had been numerically observed~\cite{PhysRevLett.95.240402,PhysRevA.77.013624,PhysRevB.75.100503,Roux_2008}. Defining a phase boundary between the charge-$4e$ and charge-$2e$ phases is challenging, but our data suggests it lies close to or at $V/U=0$.

\section{Discussion}
The attractive fermionic Hubbard model describes a system\ where a (charge-$2e$) superconducting ground state is unambiguously established theoretically. Realizing superconductivity in the attractive Hubbard model is the declared and realistic goal of cold-atom research~\cite{doi:10.1126/science.ade4245}. In our work, we have shown that the two-orbital extension of this model realizes a charge-$4e$ superconducting phase at intermediate and strong coupling in a substantial part of its phase diagram, and by considering the thermodynamic limit. In particular, we found the phase to persist for weak (attractive) inter-orbital interactions, which addresses the probably biggest limitation in realizing it with cold atoms in experiments.

In addition to the implementation in cold-atom experiments, several theoretical avenues for extending our work present themselves: The approach to construct a charge-$4e$ superconductor can be replicated in higher dimensions. Specifically in two dimensions, the physics of Mott and charge-transfer insulators recently described in van der Waals heterostructures may be exploited towards a solid-state experimental realization. Furthermore, the charge-$4e$ order parameter realized in our model,  which is to good approximation the local quartet operator $\hat{Q}$, transforms trivially under all point group (or internal) symmetries of the model (inversion, the exchange between the two orbitals, and time reversal symmetry). Thus, the type of $4e$ superconductivity found here can be denoted conventional, paralleling the classification of charge-$2e$ superconductors. Exploring unconventional charge-$4e$ superconductors, in other words charge-$4e$ order parameters that transform as a nontrivial representation of the point group, could lead to the discovery of new topological states.

\textbf{Note added} Recently, Ref.~\cite{chirolli2024cooper} appeared, which is closely related to the work presented here. There, the authors consider a similar construction as the one used here, applied in the context of quantum devices.

\begin{acknowledgments}
The authors acknowledge useful discussions with Bartholomew Andrews, Edwin Miles Stoudenmire, Johannes Motruk, Martin Zwierlein, J\"org Schmalian, Daniele Guerci, and Isidora Araya Day. The authors also acknowledge Nikita Astrakhantsev for his contributions to some of the preliminary calculations of this work.
We acknowledge the use of the software packages \texttt{QuSpin}~\cite{quspin1,quspin2} for ED calculations, \texttt{TeNPy}~\cite{tenpy} for the iDMRG numerical simulations, and \texttt{Pymablock}~\cite{Pymablock} to aid the low energy effective model derivation.
M.O.S. acknowledges funding from the Forschungskredit of the University of Zurich (Grant No. FK-23-110). T.N. acknowledges support from the Swiss National Science Foundation through a Consolidator Grant (iTQC, TMCG-2\_213805). 
\end{acknowledgments}

\section*{Data availability}
The code used to generate the data in this work and data underlying the figures is available online~\cite{data_repository}. 

\begin{appendix}\label{app}
\setcounter{equation}{0}
\renewcommand\theequation{A\arabic{equation}}
\setcounter{figure}{0}
\renewcommand\thefigure{A\arabic{figure}}

\section{Charge-2\textit{e} model}\label{app:charge2e model}
\begin{figure*}[t]
    \centering\includegraphics{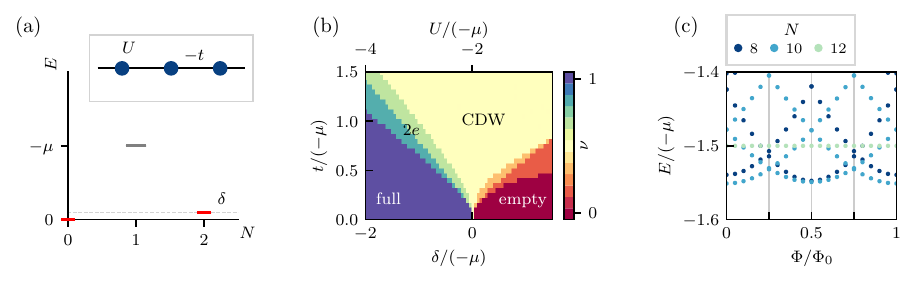}
    \caption{\textbf{Charge-2\textit{e} model} (a) Energy levels for the single site Hamiltonian, with $\delta=-2\mu+U$. The inset shows the schematic chain structure. (b) Filling of the ground state as a function of $\delta$ and $t$ as obtained from ED on a chain of length $L=6$. (c) Response of the energy spectrum under flux insertion for a chain of length $L=6$, with $\delta=-0.25$ and $t=0.3$.}
    \label{fig:SI charge2e}
\end{figure*}

In the main text, we presented a charge-$4e$ model built from a Bose-Einstein-condensate limit, Eq.~\eqref{eq:1D charge4e model}. In this section, we discuss the analogous construction for a charge-$2e$ system. 
We introduce the 2$e$ model by considering a one-dimensional (1D) chain with a single spinful orbital per site, with $\hat{c}^{\dagger}_{i, \sigma}$ ($\hat{c}_{i, \sigma}$) the creation (annihilation) operator of the electron with spin $\sigma \in \{\uparrow, \downarrow\}$ at the $i$th site.
The Hamiltonian for the charge-$2e$ model is the 1D Hubbard model,
\begin{equation}\label{eq:1D charge2e model}
\begin{split}
    \hat{H}_{2e} &= \sum_i \hat{H}_i - t \sum_{i, \sigma} \hat{c}^{\dagger}_{i, \sigma} \hat{c}_{i +1, \sigma}, \\
    \hat{H}_i &= - \mu \sum_{\sigma} \hat{n}_{i, \sigma} + U \hat{n}_{i, \uparrow} \hat{n}_{i, \downarrow},
\end{split}
\end{equation}
The Hamiltonian for an individual site, $\hat{H}_i$, has a four-dimensional Hilbert space, and the two-fold occupied state has energy $\delta\equiv(-2\mu+U)$. For $-\mu<0$, $U<0$ and $\delta \ll 1$, the low energy subspace is made of the empty and two-fold occupied state (Fig.~\ref{fig:SI charge2e}(a)). As in the main text, we express the parameters in units of $(-\mu)$ in the following, unless otherwise specified. 
The hopping term in Eq.~\eqref{eq:1D charge2e model} induces coherence between sites and leads to superfluidity in the thermodynamic limit, for the right choices of $\delta,\, t$. The ground-state filling evaluated within ED~\footnote{All the ED calculations presented in this work have been obtained with the package \texttt{Quspin} (version 0.3.6)~\cite{quspin1,quspin2}.} is shown in Fig.~\ref{fig:SI charge2e}(b), for a chain of length $L=6$. The figure shows two regions where the ground state is the empty or full lattice, a charge density wave (CDW) region, where the ground state is at half filling, and intermediate regions, where the filling interpolates between $0$ or $1$ and $0.5$. Note that in the thermodynamic limit the CDW and the $2e$ gapless phases become degenerate at half filling, and the distinction in Fig.~\ref{fig:SI charge2e}(b) is an artifact of finite size effects in ED. Figure~\ref{fig:SI charge2e}(c) shows the response of the energy spectrum under flux insertion, which has minima at $\Phi/\Phi_0 = 0,\,\frac{1}{2}$, as expected from a charge-$2e$ superconductor.

\section{Derivation of Effective spin model}\label{app:effective spin model}
In this section, we outline the derivation of the low energy effective spin-1/2 model for the charge-$4e$ model introduced in the main text, Eq.~\eqref{eq:1D charge4e model}, and we show the resulting phase diagram.

At every site, the local Hilbert space of the full fermionic model Eq.~\eqref{eq:1D charge4e model} is 16-dimensional. In the low energy effective model we only retain the two lowest-energy states, namely the vacuum and the four-fold occupied state
\begin{equation}
    \ket{0}, \quad \hat{c}^{\dagger}_{i, 1, \uparrow}\hat{c}^{\dagger}_{i, 1, \downarrow}\hat{c}^{\dagger}_{i, 2, \uparrow}\hat{c}^{\dagger}_{i, 2, \downarrow}\ket{0} = \ket{4},
\end{equation}
and integrate out all the remaining states.
In the following, we consider the limit $\delta \ll 1$, where the low energy theory is valid, and we introduce the $V/U$ ratio $r$, $V = r U $ and $U =(\delta - 2)/(1+2 r)$. Note that all the quantities are expressed in units of $-\mu$.
We take the hopping term of the Hamiltonian in~\eqref{eq:1D charge4e model} to be the perturbation, with $t\ll1$, and then expand in $\delta \ll 1$. For the perturbation theory, we consider terms up to fourth order in $t$, as this is the minimal order of the perturbative expansion that allows to describe hopping processes of quartets, and we only retain terms to first order in $\delta$.

In the following, we consider one, two and three sites processes in the fermionic model, and write their perturbation theory contributions~\footnote{We supported the analytical derivation of the model in Eqs.~\eqref{eq:spin eff model} with the python package \texttt{Pymablock} (version 0.0.1)~\cite{Pymablock}.}. There are no corrections coming from four or more than four site processes up to this order in perturbation theory.

The Hamiltonian projected on a single site does not contain corrections due to the tunneling term, but it has only the diagonal entries
\begin{equation}
    \bra{0}H\ket{0} =0, \ \bra{4}\hat{H}\ket{4}=2\delta.
\end{equation}
Two site contribution are both diagonal and off diagonal, 
\begin{equation}
\begin{split}
    \bra{4, 0} H \ket{4, 0} &= 2 \delta - \left(\delta +2\right) t^2 - \frac{\left(2 r^2-r+1\right) }{4 r (r+1)} t^4 \\
    &\qquad + O(t^5, \delta^2, t^4\delta) \\
    &= W t^2 + Y(r) t^4 + O(t^5, \delta^2, t^4\delta)
\end{split}
\end{equation}
and
\begin{equation}
    \begin{split}
    \bra{0, 4} H \ket{4, 0} &= -\frac{(2 r+1) (5 r+1) }{4 r (r+1)} t^4 \\
    &\qquad + O(t^5, \delta^2, t^4\delta) \\
    &= Z(r) t^4  + O(t^5, \delta^2, t^4\delta),
    \end{split}
\end{equation}
respectively.

Three site contributions, with the two and one site contributions subtracted, are only diagonal and of two types
\begin{equation}
\begin{split}
    h_{400}&=\bra{4, 0, 0} H \ket{4, 0, 0} - \bra{4, 0} H \ket{4, 0} \\
    &=  \bra{0, 0, 4} H \ket{0, 0, 4} - \bra{4, 0} H \ket{4, 0}  \\
     & = \bra{4, 4, 0} H \ket{4, 4, 0} - 2\delta - \bra{4, 0} H \ket{4, 0}\\  &
     = \bra{0, 4, 4} H \ket{0, 4, 4}- 2\delta - \bra{4, 0} H \ket{4, 0} \\
      &=-\frac{1}{2}t^4 + O(t^5, \delta^2, t^4\delta),
\end{split}
\end{equation}
and
\begin{equation}
\begin{split}
    h_{040}& =\bra{0, 4, 0} H \ket{0, 4, 0} + 2\delta - 2 \bra{4, 0} H \ket{4, 0} \\
    &=  \bra{4, 0, 4} H \ket{4, 0, 4} - 2 \bra{4, 0}H \ket{4, 0} \\
    &= \frac{2 r (2 r+3)}{(3 r+2) (4 r+1)} t^4 + O(t^5, \delta^2, t^4\delta) \\
    &= X(r) t^4 + O(t^5, \delta^2, t^4\delta).
    \end{split}
\end{equation}

To summarize, we have defined
\begin{equation}
    \begin{split}
        W t^2 &= -(2+\delta)t^2,\\
        Y(r) t^4 &= - \frac{\left(2 r^2-r+1\right) }{4 r (r+1)} t^4 \\
        Z(r) t^4 &= -\frac{(2 r+1) (5 r+1) }{4 r (r+1)} t^4 \\
        X(r) t^4 &=\frac{2 r (2 r+3)}{(3 r+2) (4 r+1)} t^4.
    \end{split}
\end{equation}

We first map the low-energy fermionic model to a hardcore bosonic model, and we then translate the hardcore bosonic model to a spin-1/2 model. We identify the bosonic creation and annihilation operators
\begin{equation}
    \hat{b}^{\dagger}_i \equiv \hat{c}^{\dagger}_{i, 1, \uparrow}\hat{c}^{\dagger}_{i, 1, \downarrow}\hat{c}^{\dagger}_{i, 2, \uparrow}\hat{c}^{\dagger}_{i, 2, \downarrow}, \quad \hat{b}_i = (\hat{b}^{\dagger}_i)^{\dagger}, 
\end{equation}
which satisfy the hard-core bosonic algebra
\begin{equation}
    [\hat{b}_i, \hat{b}^{\dagger}_j] = \delta_{i, j}(1 - 2 \hat{b}^{\dagger}_i \hat{b}_i), \quad [\hat{b}_i, \hat{b}_j] = 0.
\end{equation}
At every site, the two fermionic low-energy states are mapped to the bosonic states
\begin{equation}
    \ket{0} \rightarrow \ket{0}, \quad \ket{4} \rightarrow \hat{b}^{\dagger}_i \ket{0} = \ket{1}.
\end{equation}

In the bosonic language, the one and two site contributions give
\begin{equation}
\begin{split}
    H_{\text{1,2}} =& \sum_i \left[2\delta \, n_i + Z(r) \, t^4 n_i n_{i+1} \right] \\
    & + \sum_i (W t^2 + Y(r) t^4 ) \times \\
    &\qquad \qquad \times n_i [(1 - n_{i+1}) + (1 - n_{i-1})] \\
    =&  \sum_i (2\delta + 2 W t^2 + Y(r) t^4)  \, n_i \\
    &+  \sum_i Z(r) t^4  \,n_i n_{i+1} \\
    &- \sum_i 2(W t^2 + Y(r) t^4 ) \, n_i  n_{i+1}
\end{split}
\end{equation}
and three site contributions result into
\begin{equation}
\begin{split}
   H_{\text{3}} = & \sum_i (X(r) - 1) t^4 n_i\\
   &\quad + \sum_i (X(r) + 1) t^4 n_i n_{i+1}\\
   &\qquad - \sum_i 2X(r) n_i n_{i+1}.
\end{split}
\end{equation}
The total Hamiltonian is $H_{\text{H.c. bosons}}=H_{\text{1,2}}+H_{\text{3}}$.

\begin{figure*}[ht!]
    \centering
    \includegraphics{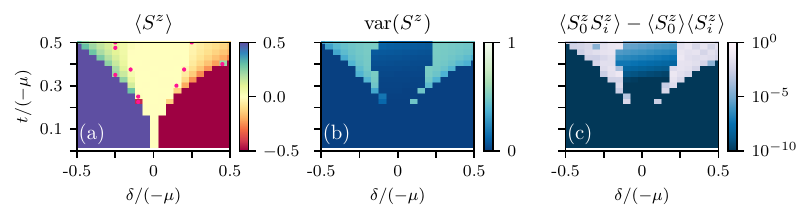}
    \caption{\textbf{Effective spin model.} (a) Expectation value of the $z$-component of the spin $S^z$, (b) its variance, and (c) correlations measured at a fixed distance $i=10$. These values are obtained within iDMRG with bond dimension $\chi=50$, and constant $V/U=0.8$. Data points marked by a pink dot in (a) correspond to iDMRG runs that reached the maximal number of sweeps before reaching the accuracy $\Delta E$.}
    \label{fig:SIFig_spin}
\end{figure*}

The hardcore bosons can be expressed in terms of spin-1/2 operators $S^k_i = \sigma^k_i/2$ ($k=x,y,z$, $\sigma^k$ Pauli matrices) acting on site $i$
\begin{equation}
    \begin{split}
    &\hat{b}^{\dagger}_i = 2 S^+_i = (S^x_i + \mathrm{i}S^y_i), \\
    &\hat{b}_i = 2 S^-_i = (S^x_i - \mathrm{i}S^y_i), \\
    &\hat{b}^{\dagger}_i \hat{b}_i = \hat{n}_i= \left(S^z_i + \frac{\mathbb{1}}{2}\right),
    \end{split}
\end{equation}
which satisfy the relations
\begin{equation}
\begin{split}
     [S^-_i, S^+_j] = -2\delta_{ij} S^z_i, \quad (\sigma^{\pm}_i)^2 =0, \\
     S^x_i S^x_{i+1}+ S^y_i S^y_{i+1} = 2 (S^+_i S^-_{i+1}+ S^-_i S^+_{i+1}).
     \end{split}
\end{equation}
In terms of spin operators, the low energy effective Hamiltonian becomes Eq.~\eqref{eq:spin eff model} in the main text, namely
\begin{subequations}\label{eq:SI eff spin model}
\begin{equation}
    \begin{split}
    H_{\mathrm{spin}} = \sum_i & \left[  J (S^x_i S^x_{i+1} + S^y_i S^y_{i+1} + \Delta S^z_i S^z_{i+1})\right] \\
    & + \sum_i \left[  h S^z_i + K S^z_i S^z_{i+2} \right],
    \end{split}
\end{equation}
with
\begin{equation}
    \begin{split}
      &J = 2 Z(r) t^4, \\
      &J \Delta = - 2 (W t^2 + Y(r) t^4 + X(r) t^4), \\
      &\Delta = -(W t^2 + Y(r) t^4 + X(r) t^4)/(Z(r) t^4), \\
      &h  = 2\delta, \quad K = (X(r)+1) t^4
      \end{split}
\end{equation}
\end{subequations}
At $r=1$ we retrieve the SU(4) symmetric case, where the effective model parameters assume the values
\begin{equation}
    \begin{split}
      &J = -\frac{9}{2} t^4, \\
      &J \Delta  = 2 t^2 (2+\delta)  -\frac{3}{10} t^4, \\
      &h  = 2\delta, \quad K = \frac{7}{5} t^4.
    \end{split}
\end{equation}
Figure~\ref{fig:SIFig_spin} shows some observables evaluated on the two-site iDMRG ground state of~\eqref{eq:SI eff spin model} as a function of $\delta,\, t$, at constant $r=0.8$.

\section{Quarter of a magnetic flux quanta periodic response}\label{app:ED flux insertion}
\begin{figure}[th]
    \centering
    \includegraphics[width=0.45\textwidth]{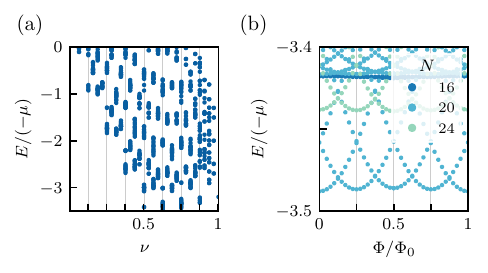}
    \caption{\textbf{Spectrum periodicity under flux insertion.} (a) ED spectrum on the $L=8$ chain with periodic boundary conditions, as a function of filling sector $\nu = N/(4L)$, with $N$ the particle number. (b) ED spectrum on the $L=8$ chain with periodic boundary conditions as a function of inserted flux $\Phi$, normalized by the flux quanta $\Phi_0$. Colors distinguish between states with different $N$, as indicated in the legend. For both panels, $\delta=-0.2$, $t=0.35$, and $V/U=0.2$. The data at $\Phi_0/\Phi_0 > 0.5$ are obtained by repeating the spectrum evaluated at $1 - \Phi_0/\Phi_0$.}
    \label{fig:ED flux}
\end{figure}
We employ ED on a finite chain to gather evidence for a phase-coherent charge-$4e$ superconducting state. The  spectrum of a $L=8$ chain with periodic boundary conditions, shown in Fig.~\ref{fig:ED flux}(a), shows the low-energy modes in the sectors with particle numbers $N=4n$, $n\in\mathbb{Z}_+$, and the absolute ground state for $N=20$ particles, which is away from half filling. We further observe a low-lying state at $\nu=1/2$ ($\nu:=N/(4L)$), which we will identify as a charge density wave (CDW) below. 

Figure~\ref{fig:ED flux}(b) depicts the lowest-energy state energy spectrum as a function of flux $\Phi$ through the ring. Conveniently, ED allows to probe the response to such a flux through modification of the tunneling amplitude $t \rightarrow e^{\mathrm{i} \varphi} t$ at every site, with $\varphi = 2\pi \Phi/(L \Phi_0)$ with $\Phi_0 = h/e$ the flux quantum.
The spectrum shows four minima with period $\pi/2$, hinting at a $\Phi_0/4$ periodicity under flux insertion, the telltale signature of the charge-$4e$ superconducting phase. (Note that we attribute the absence of exact degeneracy of states at $\varphi=0,\,\pi/2,\,\pi,\,3\pi/2$ to finite-size effects.)
Importantly, the finite curvature of the spectrum close to $\varphi=0$ is a measure for the non-zero superfluid weight $D_s$, which is indicative of a coherent state~\cite{JMKosterlitz_1973,PhysRevLett.39.1201,PhysRevLett.65.243,PhysRevB.47.7995,PhysRevLett.68.2830}. More precisely, the superfluid weight ground state is $D_s = 0.25$. For comparison, the curvature of the half-filled state is 0.007.

\section{iDMRG}\label{app:iDMRG}
\subsection{iDMRG parameters}\label{app:iDMRG parameters}
The iDMRG results presented throughout this work have been obtained by using the two-site iDMRG algorithm implemented in the \texttt{TeNPy} library (version 0.10.0)~\cite{tenpy}.
The two-site algorithm uses two sites of the Hamiltonian introduced in Eq.~\eqref{eq:1D charge4e model} to perform the iDMRG minimization.

The main parameters for the iDMRG runs used throughout this work are
\begin{itemize}
    \item \texttt{"chi\_list"}: \texttt{\{0:$\chi/2$, n\_1:$3\chi/4$, n\_2:$\chi$\}}, where $\chi$ is the bond dimension quoted with the data. The values \texttt{n\_1} and \texttt{n\_2} are of order 20 and 50, respectively.
    \item \texttt{"trunc\_params":\{"svd\_min":1.e-15, "trunc\_cut":1e-8\}}, and \texttt{"update\_env":10}.
    \item \texttt{"max\_E\_err"}=\texttt{"max\_S\_err"}=$\Delta E$. The parameter $\Delta E$, determines the two-site iDMRG convergence threshold, and we take the same for both energy and entropy convergence. Generically, $\Delta E$ is taken in the range $10^{-4}$--$10^{-8}$, depending on the phase and model.
    \item \texttt{"min\_sweeps"}: 50, and \texttt{"max\_sweeps"}: between 500 and 1000,
    \item \texttt{"mixer":True}, with parameters:
    
    \texttt{"mixer\_params":\{"amplitude":1.e-5, "decay":1.2, "disable\_after":30\}}
\end{itemize}
The remaining parameters are set to the \texttt{TeNPy} default.
The model in Eq.~\eqref{eq:1D charge4e model} in the main text is encoded in a ladder structure, see \texttt{tenpy.models.lattice.Ladder} in \texttt{TeNpy}, where the two orbitals labeled by $\ell=1,2$ are encoded as the two spinful fermionic sites at the two ends of the ladder rungs.
 
To implement spinful fermionic sites, we use the \texttt{TeNPy} structure \texttt{site.SpinHalfFermionSite}, with the options \texttt{cons$\_$N = "None", cons$\_$Sz = "None"}, which imply no local conservation of charge and spin $z$-component for the fermionic degrees of freedom, respectively. Also, the states we choose as initial ansatz for the starting point of the iDMRG algorithm mix different particle number sectors. This, with the choice of \texttt{cons$\_$N, cons$\_$Sz}, ensures that the iDMRG algorithm is not constrained to a fixed particle filling, therefore effectively realizing the ground state minimization in the grand canonical ensemble. This allows to run the calculations without knowing the ground state filling a priori. Note that the Hamiltonian in Eq.~\eqref{eq:1D charge4e model} for the charge-$4e$ model conserves particle number, and therefore the true ground state (which is obtained in the limit $\chi\rightarrow\infty$) has a well defined filling. We comment on this aspect more in detail in App.~\ref{app:correlations}.

\subsection{Phase diagram from iDMRG}\label{app:iDMRG phase diagram}
\begin{figure*}[th]
    \centering
    \includegraphics{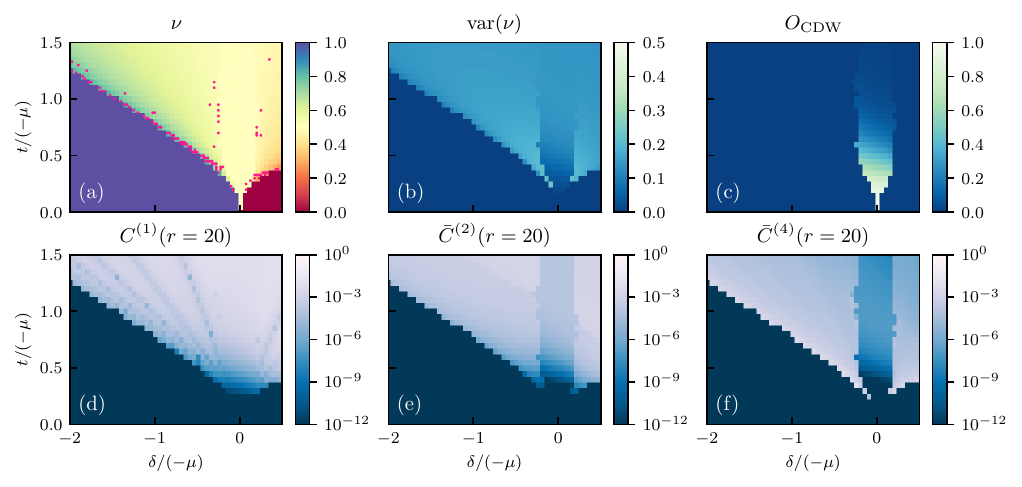}
    \caption{\textbf{Charge-4\textit{e} model.} iDMRG data evaluated at constant $V/U=0.2$, fixed bond dimension $\chi=500$, and maximal energy error $\Delta E=10^{-6}$, as a function of $t$ and $\delta$. Pink dots indicate data points for which the iDMRG run reached the maximal number of sweeps before convergence. The panels display the following quantities evaluated on the ground state: (a) Filling, (b) variance of the filling, (c) order paramter for the CDW, (d)--(f) correlations at a fixed distance $r=10$, respectively (d) $C^{(1)}$, (e) $\bar{C}^{(2)}$, and (f) $\bar{C}^{(4)}$. Note that for (d)--(f) the colorscale is logaritmic, hence correlations below the value $10^{-12}$ are flattened.} \label{fig:SIFig_iDMRGPhaseDiagram_chi_500_maxerr_1e-06}
\end{figure*}
In this section, we present in more detail the numerical results obtained within iDMRG applied to the model of Eq.~\eqref{eq:1D charge4e model} in the main text, in other words, the full fermionic model that realizes a charge-4\textit{e} superconducting phase.
To characterize the phase boundaries, we obtain a series of observables evaluated on the iDMRG ground state: the results are collected in Fig.~\ref{fig:SIFig_iDMRGPhaseDiagram_chi_500_maxerr_1e-06} at bond dimension $\chi=500$, $\Delta E =10^{-6}$.
The quantities shown in the panels of Fig.~\ref{fig:SIFig_iDMRGPhaseDiagram_chi_500_maxerr_1e-06} are defined as follows:
\begin{enumerate}
    \item Filling
    $$\nu = \sum_{i=1}^2\sum_{\ell, \sigma} \expval{\hat{n}_{i, \sigma, \ell}}_{\mathrm{GS}}/8,$$ 
    where $i$ runs over the two sites over which the iDMRG minimization is performed, and $\ell$ and $\sigma$ are the orbital and spin index respectively.
    \item Variance of the onsite filling
    $$\text{var} (\nu) = \frac{1}{16}\sum_{ \ell, \sigma} \expval{\hat{n}^2_{i, \sigma, \ell}}_{\mathrm{GS}}-\nu^2.$$
    \item The CDW order parameter
    $$O_{\text{CDW}} = \frac{1}{16} \sum_{i=1}^2 \sum_{\ell, \sigma} (-1)^{i} \expval{\hat{n}_{i, \ell, \sigma}}_{\mathrm{GS}}.$$
    \item The correlations $C^{(1)}$, $\bar{C}^{(2)}\equiv C^{(2)}_{\bar{\ell} = 0}$ and $\bar{C}^{(4)}$, defined in Eq.~\eqref{eq:corr with subtraction} in the main text.
\end{enumerate}
The jumps in the values of the filling, correlations and the CDW order parameters allow us to trace the phase boundaries sketched in Fig.~\ref{fig:charge4e model} of the main text. The transitions between empty and full lattice towards the gapless charge-$4e$ phase is sharply defined by the jump in the value of $\nu$ as well as the correlations, and the same holds for the transition between CDW and charge-$4e$ phase. The phase boundaries that separates the charge-$4e$ phase and the CDW from the Luttinger liquid are less sharp, and therefore harder to numerically pinpoint. We trace them based on the jump in $\bar{C}^{(2)}$, where the transition is more evident.

\begin{figure*}[th]
    \centering
    \includegraphics{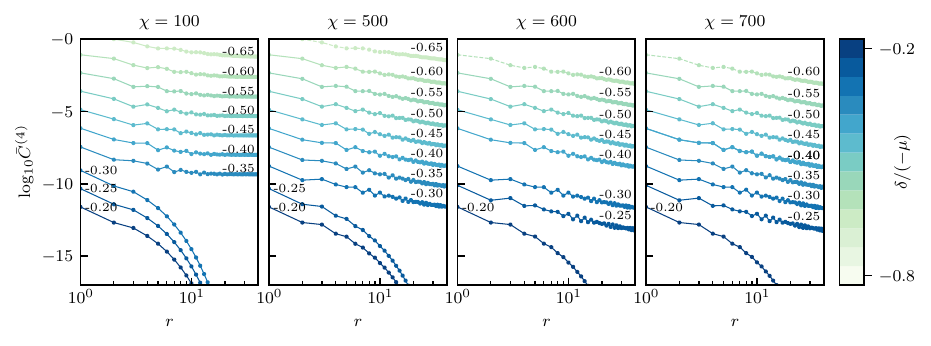}
    \caption{\textbf{Bond dimension scaling.} Log-log fountain plot of $
    \bar{C}^{(4)}(r)$ correlations for different bond dimension $\chi$ (specified by the panel titles). The correlations are evaluated within two-site iDMRG, with $\Delta E =10^{-6}$, at fixed $t/(-\mu)=0.6$ and varying $\delta/(-\mu) \in [-0.8, -0.2]$. For each value of $\delta$, correlations are shifted by an equally spaced offset, for visibility, and different values of $\delta$ are both labeled by the corresponding value and distinguished by color. In the phase with ground state filling $\nu=1$, correlations are constant and zero, therefore they do not appear in the panels. This choice of parameters drives the system across two phase boundaries: first between the full lattice ($\delta\leq-0.65$) and the $4e$ phase ($\delta\geq-0.60$), and then between $4e$ ($\delta\leq-0.30$) and the CDW ($\delta\geq-0.35$) phases. Dashed lines indicate that the iDMRG calculation reached the maximal number of sweeps before reaching the energy precision, which occurs close to the phase boundaries values.}
    \label{fig:SI_bonddimscaling}
\end{figure*}

Figure~\ref{fig:SI_bonddimscaling} shows the behavior of the $\bar{C}^{(4)}$ correlations across two phase transitions: between the fully-filled ground state and the $4e$ phases, and between the $4e$ and CDW phases, for different choices of bond dimension $\chi$. The figure suggests that the results are converged to a sufficient degree for $\chi=500$ and thus, showcases how the choice of $\chi=500$ for the phase diagram allows to trace the same boundaries that would be obtained at higher values of $\chi$, within sufficient accuracy, while being more accessible numerically.

\subsection{Luttinger liquid}\label{app:Luttinger Liquid}
\begin{figure*}[th]
    \centering
    \includegraphics{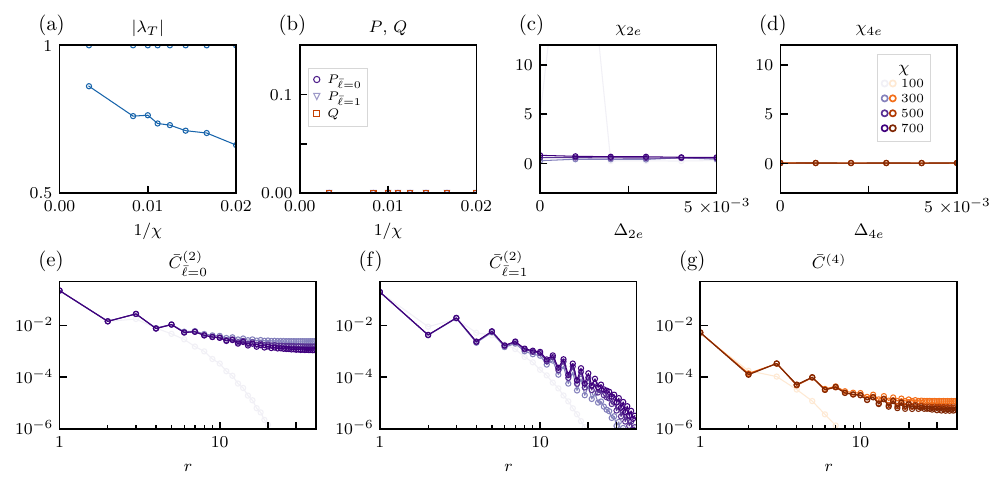}
    \caption{\textbf{iDMRG data for the Luttinger liquid.} Luttinger liquid iDMRG data evaluated at $\delta=-0.4$, $t=1.4$ and $V/U =0.2$ (and $\Delta E = 10^{-6}$ in the iDMRG parameters). (a) Transfer matrix leading and subleading eigenvalues, in absolute value, as a function of inverse bond dimension $1/\chi$. (b) Transfer matrix eigenvector overlaps $P$ and $Q$, as defined in Sec.~\ref{sec:Transfer matrix}, as a function of $1/\chi$. (c),(d) Superconducting susceptibilities (c) $\chi_{2e}$ and (d) $\chi_{4e}$ introduced in Eq.~\eqref{eq:susceptibilities} as a function of mean-field perturbation strength, $\Delta_{2e}$, and $\Delta_{4e}$, respectively. (e)--(g) Log-log plot of two- and four-body correlations defined in Eq.~\eqref{eq:corr with subtraction}, as a function of distance $r$ and for varying $\chi$. The panels show (e) $\bar{C}^{(2)}_{\bar{\ell}=0}$, (f) $\bar{C}^{(2)}_{\bar{\ell}=1}$ and (g) $\bar{C}^{(4)}$, respectively. The transparency of the data in panels (c)--(g) distinguishes different values of $\chi$, as shown in the legend in panel (d). The scale of each panel is adjusted to match the scale chosen in Figs.~\ref{fig:transfer matrix},~\ref{fig:corr and suscep}, except for panels (e)--(g), where the $y$-range is extended for visibility of panel (g).}
    \label{fig:SI Luttinger liquid}
\end{figure*}

For completeness, Fig.~\ref{fig:SI Luttinger liquid} shows the iDMRG data for a point in parameter space falling in the Luttinger liquid phase.
In panels (a)-(b), we show the data corresponding to Fig.~\ref{fig:transfer matrix} of the main text, and (c)--(g) to Fig.~\ref{fig:corr and suscep}.
As visible in Fig.~\ref{fig:SI Luttinger liquid}(e)--(g), the Luttinger liquid phase is characterized by an algebraic decay of both the $\bar{C}^{(2)}_{\bar{\ell}=0}$ and $\bar{C}^{(4)}$ correlations. In fact, the transition to the Luttinger liquid phase is most visible in the jump of the value of the $\bar{C}^{(2)}_{\bar{\ell}=0}$ correlations in Fig.~\ref{fig:SIFig_iDMRGPhaseDiagram_chi_500_maxerr_1e-06}. The transfer matrix leading and subleading eigenvalues, Fig.~\ref{fig:SI Luttinger liquid}(a), indicate that the phase is gapless, and Fig.~\ref{fig:SI Luttinger liquid}(b) shows that there is no leading transfer matrix eigenvector overlap. Finally, panels (c)--(d) of Fig.~\ref{fig:SI Luttinger liquid} indicate that this phase does not display a peak in the susceptibility to mean-field superconducting perturbations, in contrast to the case of the charge-$2e$ and charge-$4e$ phases.
Data obtained at $\chi=100$ is subject to dominant numerical fluctuations, as this bond dimension is too low to be able to capture the properties of the Luttinger liquid phase. This is clear from the exponential decay of the correlations in Figs.~\ref{fig:SI Luttinger liquid}(e), (g) at $\chi=100$, which do not occur for any other higher value of $\chi$.

\subsection{Correlations}\label{app:correlations}
In this section, we analyze in more detail some of the properties of the correlations in the charge-$4e$ system, as a complement to the data shown in Fig.~\ref{fig:corr and suscep}. First, we show the bare correlations, for completeness, and we then discuss the leveling off that the algebraically decaying correlations undergo at large distances.

\begin{figure*}[th]
    \centering
    \includegraphics{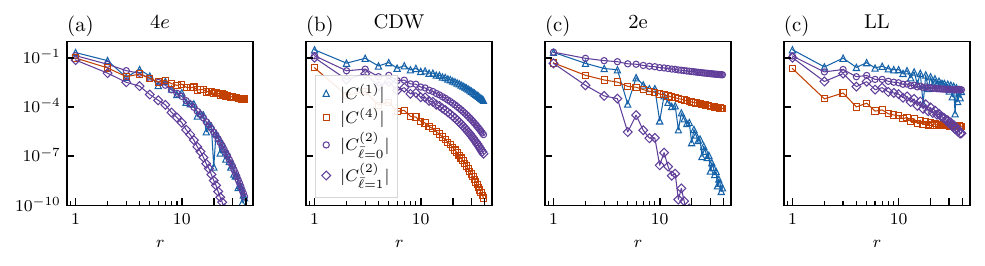}
    \caption{\textbf{Bare correlations.} Log-log plot of the bare correlations evaluated on the ground state of the model in Eq.~\eqref{eq:1D charge4e model}, for the (a) $4e$, (b) CDW, (c) $2e$, and (d) Luttinger liquid phases. The panels correspond to the same choice of model parameters of Fig.~\ref{fig:corr and suscep} in the main text, for (a)--(c), and Fig.~\ref{fig:SI Luttinger liquid} for (d): (a) $\delta=-0.5,\, t=0.6, \,V/U=0.2$, (b) $\delta=0,\, t=1, \, V/U=0.2$, (c) $\delta=-0.3, \,t=0.5, \,V/U=0$, and (d) $\delta=-0.4, \,t=1.4, \,V/U=0.2$. Here all correlations are obtained at fixed bond dimension $\chi=700$.}
\label{fig:SI_allcorrelations}
\end{figure*}

In Fig.~\ref{fig:SI_allcorrelations}, we show the bare (without subtraction) correlations, namely $C^{(1)}$, $C^{(2)}_{\bar{\ell}}$, and $C^{(4)}$, for the same choices of model parameters of Figs.~\ref{fig:corr and suscep} and~\ref{fig:SI Luttinger liquid}. The oscillations in the absolute value of the correlations are a numerical effect, and they can be suppressed for instance in two-dimensions by considering larger lattices along the added dimension~\cite{Andrews_2021}.

\begin{figure*}[th]
    \centering
    \includegraphics{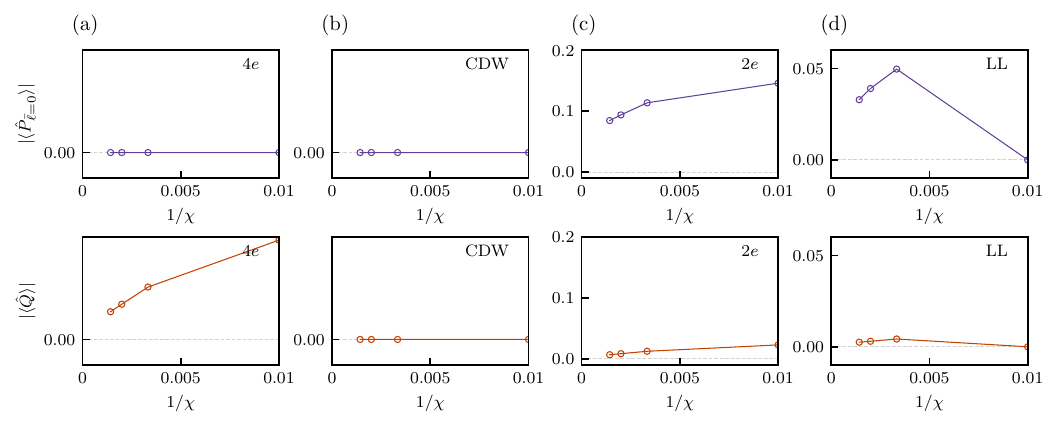}
    \caption{\textbf{Pair and quartet expectation value at finite bond dimension.} Expectation value on the iDMRG ground state of the operator (top row) $\hat{P}_{\bar{\ell}=0}$ and (bottom row) $\hat{Q}$ without mean-field perturbation ($\Delta_{2e}=0$, $\Delta_{4e}=0$) as a function of inverse bond dimension, $1/\chi$. The panels are evaluated in the (a) $4e$, (b) CDW, (c) $2e$ and (d) Luttinger liquid (LL), phases, with the same parameters as Figs.~\ref{fig:transfer matrix}, \ref{fig:corr and suscep} for (a)--(c), and Fig.~\ref{fig:SI Luttinger liquid} for (d).}
    \label{fig:SI_ExpValPQ}
\end{figure*}

\begin{figure*}[th]
    \centering
    \includegraphics{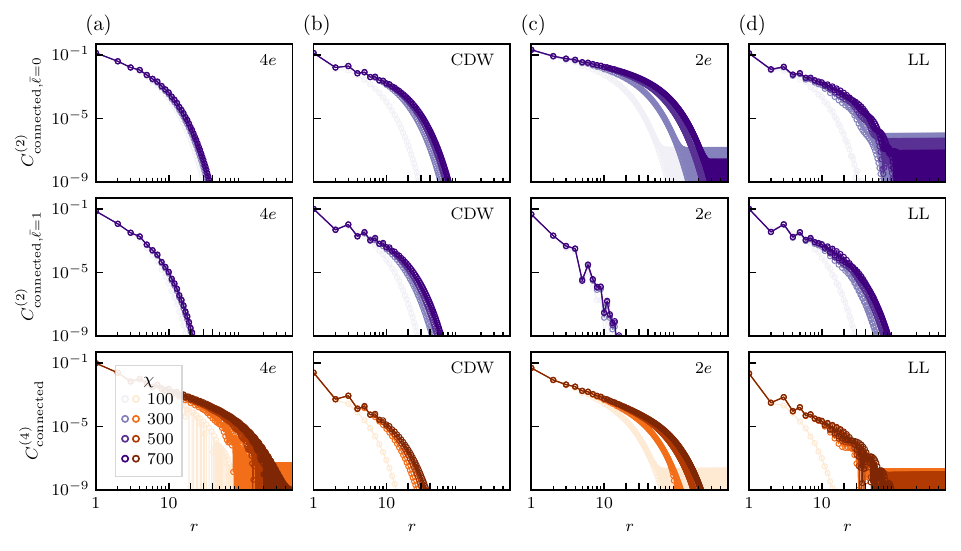}
    \caption{\textbf{Connected correlations.} Connected correlations as defined in Eqs.~\eqref{eq:connected corr}, as a function of distance $r$ and for different values of bond dimension $\chi$. The top and central rows show the connected two-body correlations for the pair operators $\hat{P}_{\bar{\ell}=0}$ and $\hat{P}_{\bar{\ell}=1}$ respectively, and the bottom row shows the four-body connected correlations. The transparency of the data corresponds to different values of the bond dimension $\chi$, as indicated in the legend in the bottom row panel (a). Panels (a)--(d) correspond to parameters falling in the (a) $4e$, (b) CDW, (c) $2e$, and (d) Luttinger liquid phases, respectively, with the same choice of model parameters as the ones of Figs.~\ref{fig:transfer matrix},~\ref{fig:corr and suscep} of the main text for (a)--(c), and of Fig.~\ref{fig:SI Luttinger liquid} for (d).} 
    \label{fig:ConnectedCorr}
\end{figure*}

As pointed out in Sec.~\ref{sec:corr and susc}, the algebraically decaying correlations in Fig.~\ref{fig:corr and suscep} level-off at a constant value, after some distance $r^*_{\chi}$, with both the distance and the constant value depending on $\chi$. The distance $r^*_{\chi}$ increases with increasing $\chi$, while the constant value reached by the correlations decreases with increasing $\chi$. This behavior can be explained by  observing that the iDMRG ground state at finite $\chi$ breaks particle number conservation, which leads to a nonzero onsite expectation value of the operators $\hat{P}_0$, $\hat{Q}$. This is an artifact of the finite bond dimension, which leads to an iDMRG ground state that mixes different particle sectors, and this effect should disappear in the limit of $\chi \rightarrow \infty$.
To clarify this point, we consider the correlations in the large $r$ limit, which become
\begin{equation}
    C^{(2)}_{\bar{\ell}}(r) = \expval{\hat{P}^{\dagger}_{0, \bar{\ell}} \hat{P}_{r, \bar{\ell}}} \xrightarrow{r\rightarrow\infty} \expval{\hat{P}^{\dagger}_{0, \bar{\ell}}} \expval{\hat{P}^{\,}_{r, \bar{\ell}}},
\end{equation}
and
\begin{equation}
    C^{(4)}(r) = \expval{\hat{Q}^{\dagger}_{0} \hat{Q}_{r}} \xrightarrow{r\rightarrow\infty} \expval{\hat{Q}^{\dagger}_{0}} \expval{\hat{Q}^{\,}_{r}}.
\end{equation}
The finite expectation values on the right hand side of the limit are responsible for the constant behavior observed at large distance.
The absolute values of $\langle\hat{P}_{\bar{\ell}=0}\rangle$ and $\langle \hat{Q}\rangle$ are shown in Fig.~\ref{fig:SI_ExpValPQ} for the $4e$, CDW, $2e$, and Luttinger liquid phases, and --when finite-- they decrease with increasing $\chi$, as expected.
In Fig.~\ref{fig:ConnectedCorr}, we show the connected two-body correlations
\begin{subequations}\label{eq:connected corr}
\begin{equation}
    C^{(2)}_{\text{connected}, \bar{\ell}}(r) = \expval{\hat{P}^{\dagger}_{0, \bar{\ell}} \hat{P}_{r, \bar{\ell}}} -\expval{\hat{P}^{\,}_{r, \bar{\ell}}},
\end{equation}
and connected four-body correlations
\begin{equation}
    C^{(4)}_{\text{connected}}(r) = \expval{\hat{Q}^{\dagger}_{0} \hat{Q}_{r}} - \expval{\hat{Q}^{\dagger}_{0}} \expval{\hat{Q}^{\,}_{r}},
\end{equation}
\end{subequations}
which exponentially decay at large distances, rather than leveling off, consistently with the above considerations.

\subsection{Particle number fluctuations}\label{app:variance N}
\begin{figure*}[th]
    \centering
    \includegraphics{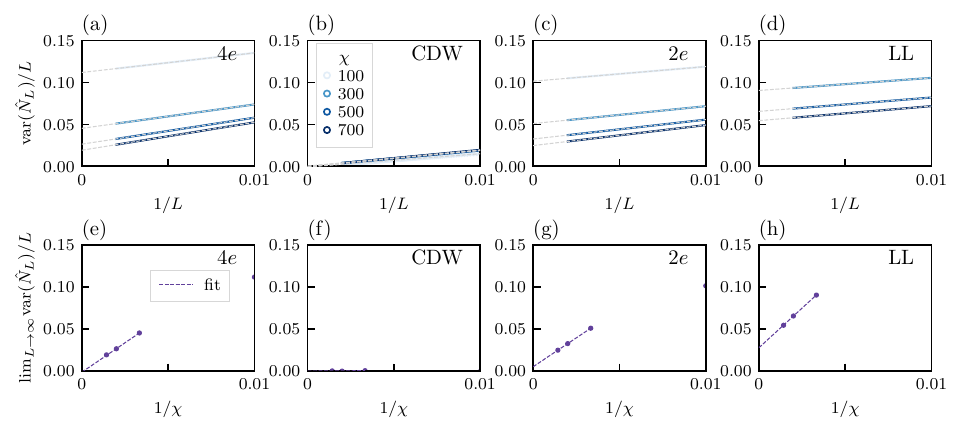}
    \caption{\textbf{Particle number fluctuations.} Top row: Value of the particle number variance for a system of size $L$ divided by $L$, as a function of $1/L$ and for different values of $\chi$. The dashed gray lines indicate the linear fit with function $a/L + b$ extracted from the data at fixed $\chi$. Bottom row: Intercept at $1/L=0$ as extracted from the top row (i.\,e. the $b$ parameter obtained from the linear fit), as a function of $1/\chi$. The dashed lines indicate the linear fit of the data points. The panels are evaluated for the (a)-(e) $4e$, (b)-(f) CDW, (c)-(g) $2e$, and (d)-(h) Luttinger liquid phases, with the same choice of model parameters as the ones in Fig.~\ref{fig:transfer matrix} and Fig.~\ref{fig:SI Luttinger liquid}.}
    \label{fig:SI_varianceN}
\end{figure*}

As a complementary analysis to the one presented in Fig.~\ref{fig:SI_ExpValPQ}, we inspect how the particle number fluctuations behave as a function of bond dimension.
To this end, we consider the particle number operator $\hat{N}=\sum_i n_i$, 
with $n_i = (n_{i, \ell=0} + n_{i, \ell=1})$ and $ n_{i, \ell} = (n_{i, \ell, \uparrow} + n_{i, \ell, \downarrow})$, and compute its variance
\begin{equation}\label{eq:variance N}
    \langle \hat{N}^2 \rangle - \langle \hat{N} \rangle^2 = \sum_{i, j} \langle n_i n_j \rangle - \left( \sum_i \langle n_i \rangle \right)^2.
\end{equation}
To evaluate Eq.~\eqref{eq:variance N}, we have to impose a cutoff on the system size, such that the summation runs from sites $0$ to a finite size $L$.
Then, we compute the variance of the particle number on a finite system of size $L$, $\hat{N}_L$, as
\allowdisplaybreaks
\begin{equation}\label{eq:variance N L}
\begin{split}
    \frac{\text{var}(\hat{N}_L)}{L}  &= \frac{1}{L} (\langle \hat{N_L}^2 \rangle - \langle \hat{N_L} \rangle^2) \\
    &= \frac{1}{L}\sum_{i=0}^L \sum_{j=0}^L \langle n_i n_j \rangle - \frac{1}{L}\left( \sum_{i=0}^L \langle n_i \rangle \right)^2.
\end{split}
\end{equation}
Panels (a)--(d) in Fig.~\ref{fig:SI_varianceN} show the value of the quantity in Eq.~\eqref{eq:variance N L} at finite $L$, and as a function of $1/L$ for different phases.
The value at $L\rightarrow\infty$ is extrapolated through a linear fit of the data at $L>100$, and it is finite at finite $\chi$. In panels (e)--(h) of Fig.~\ref{fig:SI_varianceN}, we plot the $L\rightarrow\infty$ intercept extracted from the linear fit, as a function of $1/\chi$. Note that here we exclude the data point at $\chi=100$, which is numerically less stable. The particle number fluctuations approach zero as the bond dimension increases, as expected from the fact that the Hamiltonian of the charge-$4e$ model preserves the particle number. The intercept of the $L\rightarrow\infty$ value at $\chi \rightarrow \infty$ is small but still finite, likely due to numerical inaccuracy.

\subsection{Susceptibility}\label{app:susceptibility}
\begin{figure*}[th]
    \centering
    \includegraphics{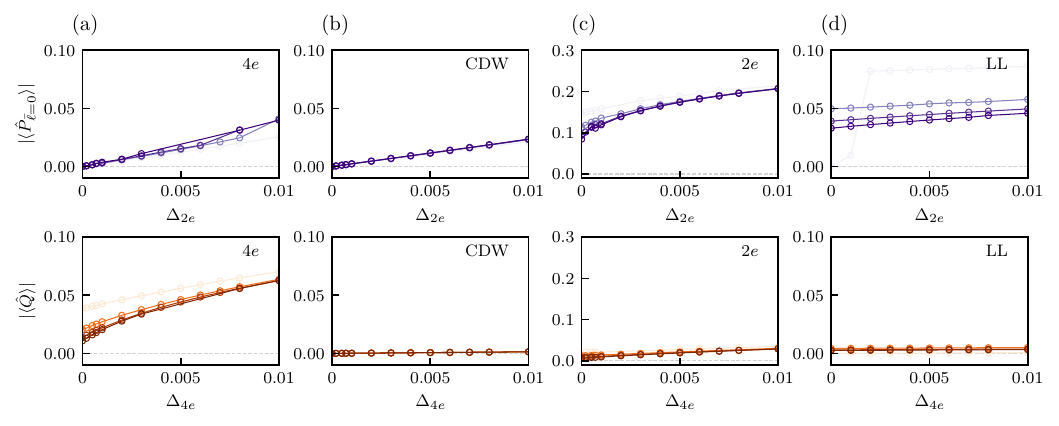}
    \caption{\textbf{Pair and quartet expectation values with mean-field perturbation.} Expectation values of (top row) $\hat{P}_{\bar{\ell}=0}$ and (bottom row) $\hat{Q}$ operators in absolute value, and as a function of perturbation strength, $\Delta_{2e}$ and $\Delta_{4e}$, respectively. The (a)--(d) panels are evaluated in the (a) $4e$, (b) CDW, (c) $2e$, and (d) Luttinger liquid phases, respectively, with the same parameter choice of Figs.~\ref{fig:transfer matrix},~\ref{fig:corr and suscep} in the main text for (a)--(c), and of Fig.~\ref{fig:SI Luttinger liquid} for (d).}
    \label{fig:SI_susceptibility}
\end{figure*}

In Eq.~\eqref{eq:susceptibilities}, in Sec.~\ref{sec:corr and susc} of the main text, we have defined the superconducting susceptibilities to charge-$2e$ and charge-$4e$ superconducting mean field perturbation.
For completeness, Fig.~\ref{fig:SI_susceptibility} shows the raw expectation values $\expval{\hat{P}_0}$ and $\expval{\hat{Q}}$ from which the susceptibilities $\chi_{2e}$ and $\chi_{4e}$, shown in Fig.\ref{fig:corr and suscep}(d),(e) are computed.
\end{appendix}
\clearpage
\bibliography{references}
\end{document}